# Lunar Near-Surface Volatile Sample Return
A White Paper submitted to the *Artemis Science Definition* call

Igor Aleinov[1,2], Michael J. Way[2], Christopher W. Hamilton[3], James W. Head[4]

[1]Columbia University, New York, NY; [2]NASA Goddard Institute for Space Studies, New York, NY; [3]University of Arizona, Tucson, AZ; [4]Brown University, Providence, RI
Contact: igor.aleinov@columbia.edu; 212-678-5587.

**The Discovery of Lunar Volatiles and the Mystery of their Origin**
Since the discovery of lunar polar volatiles in the 1990s their origin, distribution, depth and volume remain open questions. These unknowns become even more pressing with the prospect of developing a human outpost on the Moon in the near future. The volatiles are expected to be preserved in cold traps (i.e. permanently shadowed regions (PSR)), or buried beneath a layer of regolith near the poles. Typically, three sources of volatiles are considered: solar wind, volatile-rich impactors, and volcanic outgassing from the interior. The current state of knowledge does not allow us to rule out any of these sources. The information on their spatial distribution mainly comes from the neutron spectroscopy measurements performed by *Lunar Prospector* and *Lunar Reconnaissance Orbiter* missions. These observations show substantial amounts of hydrogen in the upper 1 meter of polar regolith (the detector penetration depth) and may indicate the presence of water. A single deep penetration experiment, via the *Lunar Crater Observation and Sensing Satellite* (*LCROSS*) mission, was performed in a typical PSR (Cabeus crater) and excavated the regolith up to a depth of ~4 meters (Colaprete et al. 2012). The *LCROSS* experiment showed the presence of water ~5% (by weight) plus an abundance of other volatiles, some of which, such as $H_2S$ and CO could be of volcanic origin, while others like $NH_3$, suggest a cometary source.

**Volcanic Outgassing as a Source for a Lunar Atmosphere and Near-Surface Volatiles**
The preservation of volatiles over geological timescales is defined by the competition between their production and their potential destruction in the process of regolith disruption by micrometeorite impacts ("impact gardening"), which elevates the volatiles to the more hostile environment at the surface. In their recent research, Costello et al. (2020) show that production of water by solar wind is slower than the rate of impact gardening, and thus this source is not capable of producing a significant amount of pure ice. They further suggest that any substantial pure ice deposits are likely to be the result of sudden and voluminous deposition. Such deposition could be a result of the Moon being struck by a large volatile-rich impactor (Prem et al. 2015), or due to volcanic outgassing during a major volcanic event (Needham & Kring 2017, hereafter NK2017). The patchiness of the surface ice in PSRs and the fact that it is observed mainly in very old (>3.1 Ga) craters also suggest that the volatiles could have been deposited during the peak of volcanic activity ~3.5 Ga and could be of volcanic origin (Deutsch et al. 2020). Such volcanic outgassing would produce a transient collisional lunar atmosphere. Depending on the strength and frequency of such events a millibar atmosphere could accumulate (NK2017). However, Head et al. (2020) showed that it is more likely to accumulate to a microbar scale and persist for only a few thousand years after each major eruption event. Either way, such an atmosphere would promote the delivery of volatiles from the volcanic sources to the polar regions. Aleinov et al. (2019, 2020) investigated the properties of such a hypothetical atmosphere in a "thick" limit (1–10 mb) and showed that such an atmosphere could be stable and efficiently transport volatiles from the area of volcanic outgassing at lunar maria to the polar regions.

A transient lunar atmosphere may have played an important role in the evolution of the Moon and its near-surface volatile inventory, but its precise characteristics are impossible to determine with the current state of knowledge due to the high uncertainty level in the mechanisms involved. For example, the lifetime, thickness and chemical composition of such an atmosphere would depend on the outgassing rates of different species, the atmospheric chemistry and the escape rates to space. The chemistry and the thermal escape to space would in turn depend on its climatic conditions, while suprathermal escape would be greatly affected by the activity of the Sun ~3.5 Ga (i.e., during the peak of lunar volcanic activity) and the protective properties of the Moon's ancient magnetic field, which for the period in question could have been as strong as that of modern Earth (Weiss & Tikoo 2014). None of these mechanisms are completely understood and/or well constrained, and the only reliable proxy for the ancient lunar atmosphere are the primordial volatiles deposited by it, which are expected to be preserved in the polar cold traps.

**Volatile Sample Return: A High-Priority *Artemis* Program Science Target**
The study of lunar volatiles through sample return and analysis will not only constrain the characteristics of the ancient lunar atmosphere, but also improve our understanding of the mechanisms involved, for example, the effect of increased activity of the young Sun on the Moon and the role of the Moon's magnetic field in protecting its surface.

The peak of volcanic outgassing from lunar maria corresponds to ~3.5 Ga (NK2017). It is around this time that the thickest and most persistent secondary lunar atmosphere could have formed, and most of the volcanic volatiles would be delivered to the polar cold traps. Ideally, one would collect an undisturbed sample from this period to study the ancient lunar atmosphere. With the impact gardening rate suggested by Costello et al. (2020) for a modern impactor flux ~2 meters of regolith would be reworked over the past 3.5 Ga. In an alternative scenario, with a higher impactor flux 1–3.5 Ga, the depth of reworked regolith could be as much as ~5 meters over the past 3.5 Ga. So, ideally one would want a sample from ~5 meters or deeper. Although, taking into account the uncertainty in impactor flux, a sample from any depth below 2 meters would be useful. Preferably the sample should be taken from a volatile-rich PSR, which is sufficiently old to have been present 3.5 Ga. However, a sufficiently deep sample outside of a PSR in the polar region will also be very useful, since it should contain some primordial water. If one samples a pure ice slab, a sample from a maximum possible depth is preferred, because up to 15 meters of ice could have been reworked over 3.5 Ga (Costello et al. 2020). We wish to look at the isotopic content in the sample, which in particular would allow us to distinguish between volcanic, cometary, or chondritic water. We are also seeking to determine the general abundance of volatiles such as $H_2O$, $CO_2$, $H_2S$, $CO$, $NH_3$. Some of these (like $H_2S$, $CO$) would indicate their volcanic origin, while others ($NH_3$) may point to a cometary source. The direct sampling of a PSR is preferred because some of these volatiles are unlikely to survive outside of PSRs.

We therefore concur with Jolliff et al. (2020) that lunar volatile sample return should be a high-priority within the next decade of planetary exploration and advocate that it be considered as a fundamental part of the *Artemis* program.